%% file: main.tex
\documentclass{ECOC-author}
\input{includes.tex}

\usepackage{hyperref}
\usepackage[capitalise]{cleveref}
\begin{document}

\title{EXPERIMENTAL DEMONSTRATION OF EIGHT-DIMENSIONAL MODULATION FORMATS FOR LONG-HAUL OPTICAL TRANSMISSION}

\author{Sjoerd van der Heide\ad{1}\corr, Bin Chen\ad{1,2}, Menno van den Hout\ad{1},\\Hartmut Hafermann\ad{3}, Ton Koonen\ad{1}, Alex Alvarado\ad{1}, and Chigo Okonkwo\ad{1}}

\address{\add{1}{Department of Electrical Engineering, Eindhoven University of Technology, The Netherlands.}
\add{2}{School of Computer Science and Information Engineering, Hefei University of Technology, China}
\add{3}{Mathematical and Algorithmic Sciences Lab, Paris Research centre, Huawei Technologies France SASU,
France.}
\email{s.p.v.d.heide@tue.nl}}



\begin{abstract}
    Two novel \SI{5.5}{\bit \per \fourd} modulation formats are experimentally demonstrated to transmit over \SI{9680}{\km} of \glsentryshort{SSMF}. A reach increase of \SI{29.4}{\percent} over the \SI{6}{\bit \per \fourd} \glsentryshort{PM8QAM} is shown. Furthermore, increased nonlinear tolerance with respect to \glsentryshort{PM8QAM} is shown.
\end{abstract}

\maketitle

\section{Introduction}
In optical transmission systems, the performance of a given modulation format is  determined by its tolerance
to both nonlinear interference arising from the Kerr effect, and linear accumulated \gls{ASE} noise. Therefore, designing modulation formats which increase the achievable information rate (AIR) in the presence of linear and nonlinear impairments is crucial for designing future transmission systems. 

Multidimensional constant modulus modulation formats have been shown to minimise the impact of nonlinear interference noise by minimising the signal power variations \cite{Chagnon:13,ReimerOFC2016,Kojima2017JLT,BinChenArxiv2019}.
One example of this is the \gls{4D64PRS} format we recently proposed in \cite{BinChenArxiv2019}. In \cite{Shiner:14}, \gls{8D} formats were designed to further mitigate fibre nonlinear impairments via the polarisation-balancing concept.

Previous works were only able to construct nonlinearity-tolerant \gls{8D} modulation formats in the \gls{SE} range of \SIrange{2}{4}{\bit \per \fourd}. This was achieved by set-partitioning \gls{PMBPSK} or \gls{PMQPSK} \cite{Shiner:14,El-RahmanECOC2017,Bendimerad:18}. Recently, two \gls{8D} formats with with an \gls{SE} of \SI{5.5}{\bit \per \fourd} were proposed in \cite{Bin8DPRS}. The design was based on the constant modulus and polarisation-balancing concepts. The formats were called \gls{8D2048PRS} and allow an increase in the  sensitivity and nonlinearity tolerance. Two types were proposed: Type 1 (T1) and Type 2 (T2). The formats were compared using \gls{NGMI} as performance metric, which represents the largest code rate of an ideal soft-decision \gls{FEC} in a coded modulation system with a bit-wise decoder.

In this work, the formats \gls{8D2048PRST1} and \gls{8D2048PRST2} are experimentally compared to \SI{5.5}{\bit \per \fourd} \gls{TH4D2A8PSK} \cite{KojimaOFC2017} and \SI{6}{\bit \per \fourd} \gls{PM8QAM}. A transmission distance of \SI{9680}{\km} is reported for both \gls{8D2048PRS} modulation formats, showing a reach increase of \SI{4.9}{\percent} (\SI{450}{\km}) over \gls{TH4D2A8PSK} and \SI{29.4}{\percent} (\SI{2200}{\km}) over \gls{PM8QAM}. Furthermore, \gls{8D2048PRST2} is shown to be more resilient against nonlinearities than \gls{PM8QAM}. The achieved performance is in agreement with simulation results and thus confirms the potential of these modulation formats in long-haul optical fibre transmission systems.


\section{Time-slotted \gls{8D} Modulation Format}
\label{sec:timeslot}

The 8 dimensions are obtained by using two consecutive time slots in two polarisations. \gls{8D2048PRS} is based on two consecutive \gls{4D64PRS} symbols and carries 11~bits per \gls{8D} symbol. To go from the 12 bits needed to index two consecutive \gls{4D64PRS} formats ($b_1, b_2, \ldots, b_{12}$) to the 11~bits in \gls{8D2048PRS}, the last bit is used as overhead (parity) bit \cite{BinChenArxiv2019}. In particular, $\overline{b}_{12}= {b_1\oplus b_2\oplus b_3 \oplus b_4\oplus b_5\oplus b_6\oplus b_7\oplus b_8\oplus b_9\oplus b_{10}\oplus b_{11}}$ for \gls{8D2048PRST1} and $\overline{b}_{12}={b_3\oplus b_6\oplus b_9}$ for \gls{8D2048PRST2}, where $\oplus$ and $\overline{b}$ denote
modulo-2 addition and negation, respectively.

Both \gls{8D2048PRS} formats are designed to avoid polarisation-identical symbols in both timeslots and therefore reduce the effects of cross-polarisation modulation. \Glspl{SOP} and \gls{ED} are taken into account for the selection of 2048 \gls{8D} symbols.
\gls{8D2048PRST1} is designed to use the parity bit to protect all the information bits, and thus, it leads to a higher minimum \gls{ED} and a better performance at higher \gls{SNR}. On the other hand, \gls{8D2048PRST2} is designed to be better at the lower \gls{SNR} regime since only the  least  significant bits are protected.

\Gls{SSFM} simulations with a step-size of \SI{0.1}{\km} were performed to compare the modulation formats and predict system performance.
The simulation parameters are given in Table \ref{tab:syspar} for the optical multi-span fibre link under consideration, which comprises multiple \gls{SSMF} spans, amplified at the end of each span by an \gls{EDFA}. The encoded bits are mapped according to four modulation formats: \gls{PM8QAM} (\SI{6}{\bit \per \fourd}), \gls{TH4D2A8PSK} (\SI{5.5}{\bit \per \fourd})  and two \gls{8D2048PRS} types (\SI{5.5}{\bit \per \fourd}). \Gls{TH4D2A8PSK} is generated by combining \gls{5B4D2A8PSK} and \gls{6B4D2A8PSK} from \cite{KojimaOFC2017} with a 1:1 ratio in a time domain hybrid fashion, resulting in a \gls{TH4D} modulation format. Each \gls{DWDM} channel carries independent data, where all of them are assumed to have the same transmitted power.
At the receiver, an ideal receiver is used for detection and chromatic dispersion is digitally compensated for.

\begin{figure}[t]
    \centering
    \includegraphics[width=\linewidth]{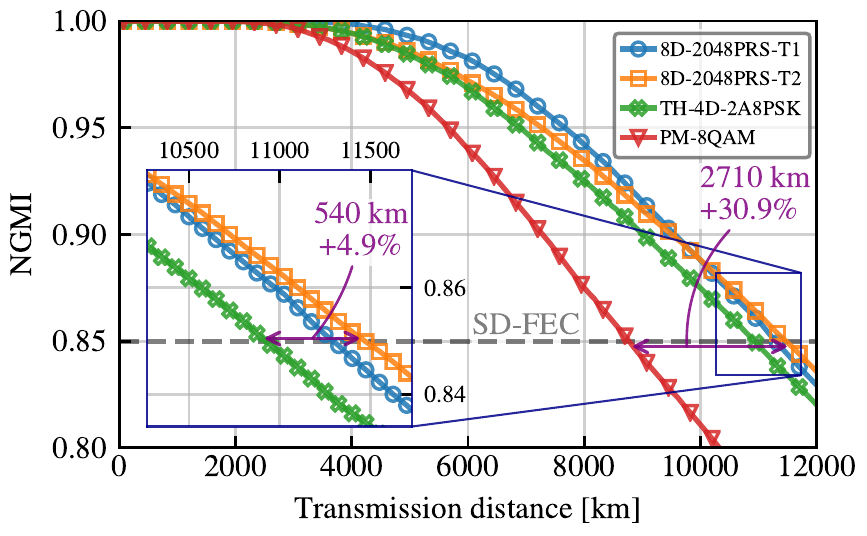}
    \caption{Simulation results: \glsentryshort{NGMI} as a function of transmission distance for the centre channel. Reach increases of 4.9\% and 30.9\% are observed for \glsentryshort{8D2048PRS} with respect to \glsentryshort{TH4D2A8PSK} and \glsentryshort{PM8QAM}, respectively.}
    \label{fig:DTsim}
\end{figure}

\begin{figure*}[b]
    \centering
    \includegraphics[width=\linewidth]{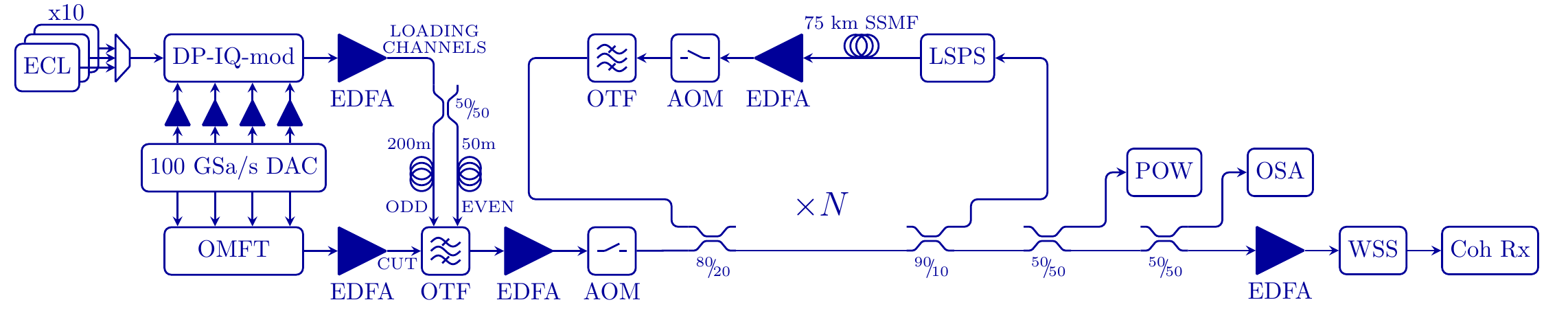}
    \caption{Experimental optical recirculating loop setup. }
    \label{fig:ECOC19_setup}
\end{figure*}

\cref{fig:DTsim} shows \gls{NGMI} \cite{AlvaradoJLT2015} as a function of the transmission distance. The results show that the \gls{8D2048PRS} formats offer large reach increases with respect to \gls{PM8QAM}. The inset in \cref{fig:DTsim} shows that \gls{8D2048PRST2} offers a \SI{4.9}{\percent} reach increase relative to \gls{TH4D2A8PSK}.

\cref{fig:DTsim} also shows that both \gls{8D2048PRST1} and \gls{8D2048PRST2} yield a \SI{30.9}{\percent} reach increase relative to \gls{PM8QAM} at a \gls{NGMI} of 0.85. 

\input{table.tex}

\section{Experimental Setup}
\cref{fig:ECOC19_setup} depicts the experimental recirculating loop setup which approximates the simulation setup of \cref{sec:timeslot}. Sequences of 2\textsuperscript{16} 4D-symbols are used for \gls{PM8QAM}. For the \gls{8D} and \gls{TH4D} modulation formats, sequences of 2\textsuperscript{15} 8D-symbols are time-multiplexed into a four-dimensional sequence of length 2\textsuperscript{16}. The generated sequence is pulse-shaped using a \gls{RRC} filter with 1\% roll-off at \SI{41.79}{\giga \baud} and uploaded to a \SI{100}{\giga\sample\per\s} \gls{DAC}. 

The \SI{1550.116}{\nm} \gls{CUT} is modulated using an \gls{OMFT}, which consists of a \gls{ECL}, a \gls{DPIQ}, an \gls{ABC} and RF-amplifiers. The multiplexed outputs of 10 \glspl{ECL} are modulated using a \gls{DPIQ}, amplified, split into odd and even, decorrelated by \SI{10200}{symbols} (\SI{50}{\m}) and \SI{40800}{symbols} (\SI{200}{\m}) with respect to the \gls{CUT}, respectively. The \gls{CUT}, odd, and even channels are combined onto the \SI{50}{\GHz} spaced \gls{DWDM} grid using an \gls{OTF}. Using \glspl{AOM}, the signal is circulated in a loop consisting of a \gls{LSPS}, a \SI{75}{\km} span of \gls{SSMF}, an \gls{EDFA}, and an \gls{OTF} used for gain equalisation. The launch power into the fibre is carefully controlled and the power per \SI{50}{\GHz} channel is equalised. The aggregate launch power of the 11 channels is denoted as 'launch power' throughout this work. 

After transmission, the signal is amplified, the \gls{CUT} selected using a \gls{WSS}, detected using an intradyne coherent receiver, and digitised by an \SI{80}{\giga\sample\per\s} real-time oscilloscope. Receiver DSP consisting of front-end compensation, \gls{CD} compensation, frequency-offset compensation, and decision-directed equalisation with in-loop phase search is performed offline. When an \gls{8D} modulation format is used, the four-dimensional sequence is time demultiplexed into \gls{8D} again. \gls{NGMI} is evaluated for the \gls{CUT} over approximately 3.5 million 8D-symbols (\gls{8D} and \gls{TH4D} formats) or 7 million 4D-symbols (\gls{PM8QAM}).

\section{Experimental Results}
\begin{figure}[t]
    \centering
    \includegraphics[width=\linewidth]{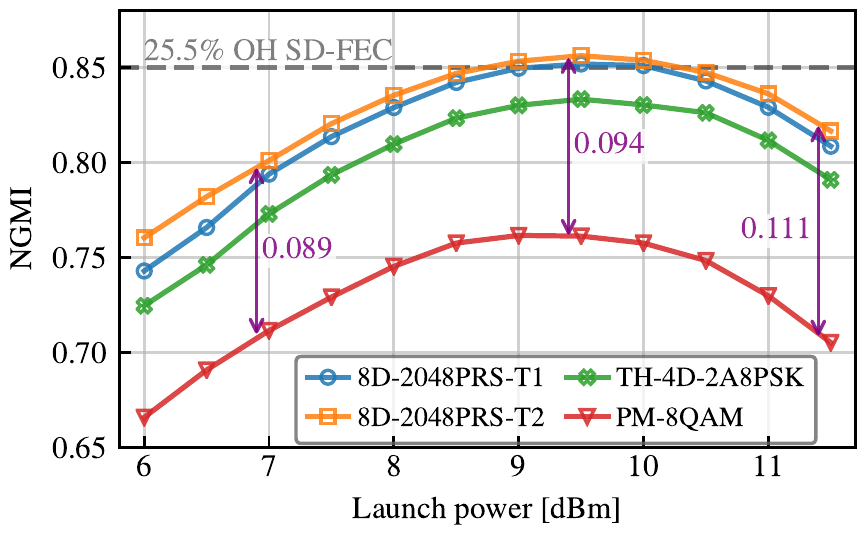}
    \caption{Experimental results: \gls{NGMI}  versus total launch power after \SI{9435}{\km}. Larger \gls{NGMI} gains at higher launch powers indicate better nonlinearity tolerance of \gls{8D2048PRS}.}
    \label{fig:LPmod}
\end{figure}

\cref{fig:LPmod} shows \gls{NGMI} versus launch power and indicates that the optimum launch power is \SI{9.5}{\dBm}. This optimum value is used in subsequent measurements. A \gls{NGMI} gain of 0.094 for \gls{8D2048PRST2} over \gls{PM8QAM} is achieved at the optimum launch power. In the linear (low launch power) regime, a similar \gls{NGMI} gain (0.089) is shown. At powers above the optimum launch power, a larger \gls{NGMI} gain of 0.111 is shown. This confirms the observation in \cite{Bin8DPRS} that the \gls{8D2048PRST2} format is more resilient against nonlinearities than \gls{PM8QAM}.

\begin{figure}[t]
    \centering
    \includegraphics[width=\linewidth]{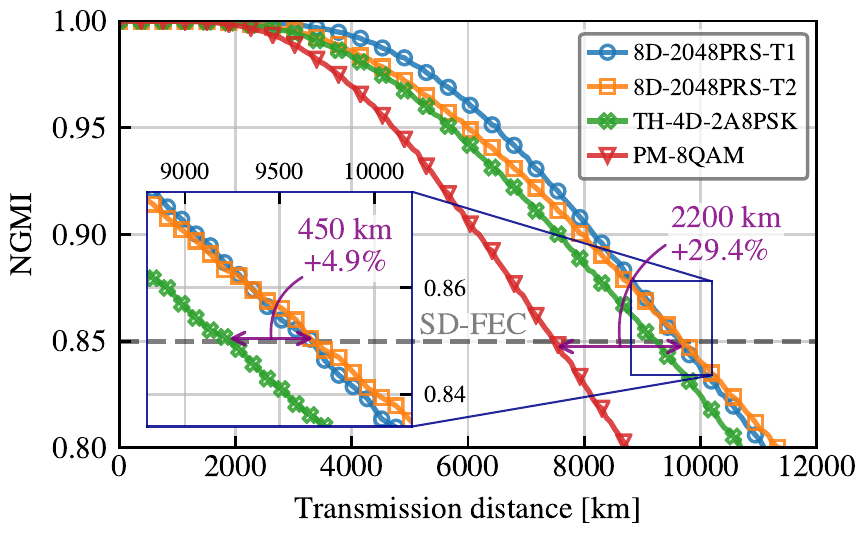}
    \caption{Experimental results: \gls{NGMI} versus transmission distance for the centre channel at a launch power of \SI{9.5}{dBm}. Both \gls{8D2048PRS} reach \SI{9680}{\km} and show a reach increase of \SI{4.9}{\percent} and \SI{29.4}{\percent} over \gls{TH4D2A8PSK} \cite{KojimaOFC2017} and \gls{PM8QAM}, respectively.}
    \label{fig:DTmod}
 \end{figure}

\cref{fig:DTmod} shows the measured \gls{NGMI} for various transmission distances. As expected from the simulations, all \SI{5.5}{\bit \per \fourd} formats perform much better than \gls{PM8QAM}. Throughout this work a \SI{25.5}{\percent} overhead \gls{FEC} is assumed with a threshold of 0.85~\gls{NGMI}. The assumed \gls{FEC} is based on a spatially-coupled \gls{LDPC} \cite{scfec} and the threshold is derived in \cite{Kojima2017JLT}. \gls{8D2048PRST1} and \gls{8D2048PRST2}, similar modulation formats but optimised for different \gls{SNR} regimes, cross at this \gls{FEC} threshold of \gls{NGMI} 0.85. This observed crossing in the experimental result is in good agreement with the simulation results of \cref{fig:DTsim}. Both \gls{8D2048PRS} formats are able to be transmitted over \SI{9680}{\km}, a \SI{4.9}{\percent} (\SI{450}{\km}) reach increase over \gls{TH4D2A8PSK}. The reach increase of both \gls{8D2048PRS} formats over \gls{PM8QAM} is \SI{29.4}{\percent} (\SI{2200}{\km}), which is very well matched to the simulation prediction of \SI{30.9}{\percent}. These experimental results confirm the performance of the novel eight-dimensional modulation formats obtained in simulation.

\section{Conclusions}
We experimentally demonstrate the transmission of two novel \SI{5.5}{\bit \per \fourd} eight-dimensional modulation formats over \SI{9680}{\km} of \gls{SSMF}. A reach increase of \SI{29.4}{\percent} over the \SI{6}{\bit \per \fourd} \gls{PM8QAM} is shown, which gives a system designer the interesting trade-off between \SI{8.3}{\percent} rate loss and \SI{29.4}{\percent} reach increase.
Furthermore, compared to the \gls{TH4D2A8PSK} format, which has the same \gls{SE} of \SI{5.5}{\bit \per \fourd}, \SI{4.9}{\percent} reach increase is shown. Note that \gls{8D2048PRS} is shown to be more resilient against nonlinearities with respect to \gls{PM8QAM}. Experimental results are in good agreement with simulation and thus confirms the potential benefits in employing these novel modulation formats in long-haul optical fibre transmission.  

\section{Acknowledgements}
Partial funding from the Dutch NWO Gravitation Program on Research centre for Integrated Nanophotonics (Grant Number 024.002.033). This research is supported in part by Huawei France through the NLCAP project. 
The work of B. Chen is partially supported by the National Natural Science Foundation of China (NSFC) under Grant 61701155.
The work of A. Alvarado is supported by the Netherlands Organisation for Scientific Research (NWO) via the VIDI Grant ICONIC (project number 15685). 
Fraunhofer HHI and ID Photonics are gratefully acknowledged for providing their Optical-Multi-Format Transmitter.

\newpage
\section{References}
\bibliographystyle{ECOC} 
\bibliography{example-ref.bib,referencesBin.bib}

\end{document}

%% file: includes.tex
\usepackage[utf8]{inputenc}
\usepackage{glossaries}
\usepackage[UKenglish]{babel}

\newcommand{\SetCapsType}{normalcaps}
\input{acronyms.tex} 

\usepackage{xcolor}
\usepackage{float}
\glsunset{8D2048PRST1}
\glsunset{8D2048PRST2}
\glsunset{5B4D2A8PSK}
\glsunset{6B4D2A8PSK}
\glsunset{4D}

\AtEndPreamble{\DeclareSymbolFont{operators}{OT1}{\rmdefault}{m}{n}}
\usepackage[per-mode=symbol]{siunitx}
\DeclareSIUnit{\sample}{Sa}
\DeclareSIUnit{\baud}{Bd}
\DeclareSIUnit{\bit}{bit}
\DeclareSIUnit{\fourd}{4D}
\DeclareSIUnit{\eightd}{8D}
\DeclareSIUnit{\dBm}{dBm}

%% file: acronyms.tex
\makeatletter
\providecommand{\SetCapsType}{smallcaps}

\long\def\@scTrue{smallcaps}
\long\def\@scFalse{normalcaps}
\newcommand{\acroSCaps}[1]{%
 \begingroup
  \ifx\SetCapsType\@scTrue 
    \textsc{#1}%
  \else
    \MakeUppercase{#1}%
  \fi
  \endgroup
}
\makeatother

\newcommand{\nAcronym}[4][]{%
	\newacronym[#1]{#2}{#3}{#4}
}

\makeatletter
\@ifpackageloaded{babel}{
    \newcommand{\usuk}[2]{%
        \iflanguage{USenglish}{#1}{#2}
    }
}{
    \newcommand{\usuk}[2]{%
        #1
    }
}
\makeatother

\usepackage[shortcuts]{extdash}
\nAcronym{4D}{\acroSCaps{4d}}{four-dimensional}
\nAcronym{4D64PRS}{\acroSCaps{4d-64prs}}{\usuk{four-dimensional 64-ary polarization-ring-switching}{four-dimensional 64-ary polarisation-ring-switching}}

\nAcronym{5B4D2A8PSK}{\acroSCaps{5b4d-2a8psk}}{5-bit four-dimensional two-amplitude 8-ary phase-shift keying}

\nAcronym{6B4D2A8PSK}{\acroSCaps{6b4d-2a8psk}}{6-bit four-dimensional two-amplitude 8-ary phase-shift keying}

\nAcronym{8D}{\acroSCaps{8d}}{eight-dimensional}
\nAcronym{8D2048PRS}{\acroSCaps{8d-2048prs}}{eight-dimensional 2048-ary polarization-ring-switching}
\nAcronym{8D2048PRST1}{\acroSCaps{8d-2048prs-t1}}{eight-dimensional 2048-ary polarization-ring-switching type 1}
\nAcronym{8D2048PRST2}{\acroSCaps{8d-2048prs-t2}}{eight-dimensional 2048-ary polarization-ring-switching type 2}

\nAcronym{ABC}{\acroSCaps{abc}}{automatic bias controller}
\nAcronym{ADC}{\acroSCaps{adc}}{analog-to-digital converter}
\nAcronym{AIR}{\acroSCaps{air}}{achievable information rate}
\nAcronym{AOM}{\acroSCaps{aom}}{acoustic optical modulator}
\nAcronym{AR}{\acroSCaps{ar}}{achievable rate}
\nAcronym{ASE}{\acroSCaps{ase}}{amplified spontaneous emission}
\nAcronym{AWGN}{\acroSCaps{awgn}}{additive white Gaussian noise}

\nAcronym{BER}{\acroSCaps{ber}}{bit error rate}
\nAcronym{BICM}{\acroSCaps{bicm}}{bit-interleaved coded modulation}
\nAcronym{BPD}{\acroSCaps{bpd}}{balanced photo-diode}
\nAcronym{BPS}{\acroSCaps{bps}}{blind phase search}

\nAcronym{CCDM}{\acroSCaps{ccdm}}{constant composition distribution matching}
\nAcronym{CD}{\acroSCaps{cd}}{chromatic dispersion}
\nAcronym{CIR}{\acroSCaps{cir}}{channel impulse response}
\nAcronym{ChUT}{\acroSCaps{chut}}{channel under test}
\nAcronym{CUT}{\acroSCaps{cut}}{channel under test}
\nAcronym{CPE}{\acroSCaps{cpe}}{carrier phase estimation}

\nAcronym{DAC}{\acroSCaps{dac}}{digital-to-analog converter}
\nAcronym{DCF}{\acroSCaps{dcf}}{\usuk{dispersion compensated fiber}{dispersion compensated fibre}}
\nAcronym{DFB}{\acroSCaps{dfb}}{distributed feedback}
\nAcronym{DGD}{\acroSCaps{dgd}}{differential group delay}
\nAcronym{DM}{\acroSCaps{dm}}{distribution matcher}
\nAcronym{DMGD}{\acroSCaps{dmgd}}{differential mode group delay}
\nAcronym{DP}{\acroSCaps{dp}}{\usuk{dual-polarization}{dual-polarisation}}
\nAcronym{DPC}{\acroSCaps{dpc}}{digital pre-compensation}
\nAcronym{DPE}{\acroSCaps{dpe}}{digital pre-emphasis}
\nAcronym{DPIQ}{\acroSCaps{dp-iqm}}{\usuk{dual-polarization IQ-modulator}{dual-polarisation IQ-modulator}}
\nAcronym{DRE}{\acroSCaps{dre}}{digital resolution enhancer}
\nAcronym{DSO}{\acroSCaps{dso}}{digital sampling oscilloscope}
\nAcronym{DSP}{\acroSCaps{dsp}}{digital signal processing}
\nAcronym{DUT}{\acroSCaps{dut}}{device under test}
\nAcronym{DWDM}{\acroSCaps{dwdm}}{dense wavelength-division multiplexing}

\nAcronym{ECL}{\acroSCaps{ecl}}{external cavity laser}
\nAcronym{ED}{\acroSCaps{ed}}{Eucledian distance}
\nAcronym{EDFA}{\acroSCaps{edfa}}{\usuk{erbium doped fiber amplifier}{erbium doped fibre amplifier}}
\nAcronym{ESS}{\acroSCaps{ess}}{enumerative sphere shaping}

\nAcronym{FDE}{\acroSCaps{fde}}{\usuk{frequency domain equalizer}{frequency domain equaliser}}
\nAcronym{FEC}{\acroSCaps{fec}}{forward error correction}
\nAcronym{FFT}{\acroSCaps{fft}}{fast Fourier transform}
\nAcronym{FMF}{\acroSCaps{fmf}}{\usuk{few-mode fiber}{few-mode fibre}}
\nAcronym[plural=FM-MCF, firstplural=\usuk{few-mode multi-core fibers}{few-mode multi-core fibres}]{FM-MCF}{\acroSCaps{fm-mcf}}{\usuk{few-mode multi-core fiber}{few-mode multi-core fibre}}

\nAcronym{GFF}{\acroSCaps{gff}}{gain flattening filter}
\nAcronym{GMI}{\acroSCaps{gmi}}{\usuk{generalized mutual information}{generalised mutual information}}
\nAcronym{GS}{\acroSCaps{gs}}{geometric shaping}

\nAcronym{HDFEC}{\acroSCaps{hd-fec}}{hard decision forward error correction}

\nAcronym[plural=IL, firstplural=insertion losses (\textsc{il})]{IL}{\acroSCaps{il}}{insertion loss}
\nAcronym{IFFT}{\acroSCaps{ifft}}{inverse fast Fourier transform}

\nAcronym{KK}{\acroSCaps{kk}}{Kramers-Kronig}

\nAcronym{LDPC}{\acroSCaps{ldpc}}{low-density parity-check}
\nAcronym{LEAF}{\acroSCaps{leaf}}{\usuk{large effective area fiber}{large effective area fibre}}
\nAcronym{LMS}{\acroSCaps{lmf}}{least means square}
\nAcronym{LLR}{\acroSCaps{llr}}{log-likelihood ratio}
\nAcronym{LO}{\acroSCaps{lo}}{local oscillator}
\nAcronym{LP}{\acroSCaps{lp}}{\usuk{linear polarized}{linear polarised}}
\nAcronym{LSPS}{\acroSCaps{lsps}}{\usuk{loop-synchronized polarization scrambler}{loop-synchronised polarisation scrambler}}

\nAcronym{MB}{\acroSCaps{mb}}{Maxwell-Bolzmann}
\nAcronym{MCF}{\acroSCaps{mcf}}{\usuk{multi-core fiber}{multi-core fibre}}
\nAcronym{MDG}{\acroSCaps{mdg}}{mode dependent gain}
\nAcronym[plural=MDL, firstplural=mode dependent losses (\textsc{mdl})]{MDL}{\acroSCaps{mdl}}{mode dependent loss}
\nAcronym{MDM}{\acroSCaps{mdm}}{mode division multiplexing}
\nAcronym{MF}{\acroSCaps{mf}}{matched filter}
\nAcronym{MI}{\acroSCaps{mi}}{mutual information}
\nAcronym{MIMO}{\acroSCaps{mimo}}{multiple-input multiple-output}
\nAcronym{MMA}{\acroSCaps{mma}}{multi-modulus algorithm}
\nAcronym{MMF}{\acroSCaps{mmf}}{\usuk{multi-mode fiber}{multi-mode fibre}}
\nAcronym{MMSE}{\acroSCaps{mmse}}{minimum mean squared error}
\nAcronym{MSE}{\acroSCaps{mse}}{mean squared error}
\nAcronym{MUX}{\acroSCaps{mux}}{multiplexer}
\nAcronym{MZM}{\acroSCaps{mzm}}{Mach-Zehnder modulator}

\nAcronym{NGMI}{\acroSCaps{ngmi}}{\usuk{normalized generalized mutual information}{normalised generalised mutual information}}
\nAcronym{NIR}{\acroSCaps{nir}}{near infra-red}

\nAcronym{OFDR}{\acroSCaps{ofdr}}{optical frequency-domain reflectometer}
\nAcronym{OMFT}{\acroSCaps{omft}}{optical-multi-format transmitter}
\nAcronym{OSA}{\acroSCaps{osa}}{\usuk{optical spectrum analyzer}{optical spectrum analyser}}
\nAcronym{OSNR}{\acroSCaps{osnr}}{optical signal-to-noise ratio}
\nAcronym{OTDR}{\acroSCaps{otdr}}{optical time-domain reflectometer}
\nAcronym{OTF}{\acroSCaps{otf}}{optical tunable filter}
\nAcronym{OVNA}{\acroSCaps{ovna}}{\usuk{optical vector network analyzer}{optical vector network analyser}}

\nAcronym{PAM}{\acroSCaps{pam}}{pulse-amplitude modulation}
\nAcronym{PAS}{\acroSCaps{pas}}{probabilistic amplitude shaping}
\nAcronym{PAPR}{\acroSCaps{papr}}{peak-to-avarage power ratio}
\nAcronym{PBS}{\acroSCaps{pbs}}{polarization beam splitter}
\nAcronym{PCVD}{\acroSCaps{pcvd}}{plasma chemical vapor depostion}
\nAcronym{PDL}{\acroSCaps{pdl}}{polarization dependent loss}
\nAcronym{PL}{\acroSCaps{pl}}{photonic lantern}
\nAcronym{PMBPSK}{\acroSCaps{pm-bpsk}}{polarization-multiplexed binary phase-shift-keying}
\nAcronym{PMQPSK}{\acroSCaps{pm-qpsk}}{polarization-multiplexed quaternary phase-shift-keying}
\nAcronym{PM8QAM}{\acroSCaps{pm-8qam}}{polarization-multiplexed 8-ary quadrature amplitude modulation}
\nAcronym{PMD}{\acroSCaps{pmd}}{polarization mode dispersion}
\nAcronym{PNOB}{\acroSCaps{pnob}}{physical number of bits}
\nAcronym{PRBS}{\acroSCaps{prbs}}{pseudorandom bit sequence}
\nAcronym{PS}{\acroSCaps{ps}}{probabilistic shaping}

\nAcronym{QAM}{\acroSCaps{qam}}{quadrature amplitude modulation}
\nAcronym{QPSK}{\acroSCaps{qpsk}}{quadrature phase shift keying}

\nAcronym{RRC}{\acroSCaps{rrc}}{root-raised-cosine}

\nAcronym{SamPerSym}{\acroSCaps{sps}}{samples per symbol}
\nAcronym{SDFEC}{\acroSCaps{sd-fec}}{soft decision forward error correction}
\nAcronym{SDM}{\acroSCaps{sdm}}{space-division multiplexing}
\nAcronym{SE}{\acroSCaps{se}}{spectral efficiency}
\nAcronym{SSFM}{\acroSCaps{ssfm}}{split-step Fourier method}
\nAcronym{SMF}{\acroSCaps{smf}}{\usuk{single-mode fiber}{single-mode fibre}}
\nAcronym{SNR}{\acroSCaps{snr}}{signal-to-noise ratio}
\nAcronym[firstplural=\usuk{states of polarization (\acroSCaps{sop})}{states of polarisation (\acroSCaps{sop})}]{SOP}{\acroSCaps{sop}}{\usuk{state of polarization}{state of polarisation}}
\nAcronym{SPS}{\acroSCaps{sps}}{\usuk{synchronized polarization scrambler}{synchronised polarisation scrambler}}
\nAcronym{SSMF}{\acroSCaps{ssmf}}{\usuk{standard single-mode fiber}{standard single-mode fibre}}
\nAcronym{SVD}{\acroSCaps{svd}}{singular value decomposition}
\nAcronym{SWI}{\acroSCaps{swi}}{swept wavelength interferometry}

\nAcronym{TDE}{\acroSCaps{tde}}{\usuk{time domain equalizer}{time domain equaliser}}
\nAcronym{TDMSDM}{\acroSCaps{tdm-sdm}}{time-domain multiplexed space-division multiplexing}
\nAcronym{TH4D}{\acroSCaps{th-4d}}{time domain hybrid four-dimensional}
\nAcronym{TH4D2A8PSK}{\acroSCaps{th-4d-2a8psk}}{time domain hybrid four-dimensional two-amplitude eight-phase shift keying}

\nAcronym{VOA}{\acroSCaps{voa}}{variable optical attenuator}

\nAcronym{WDM}{\acroSCaps{wdm}}{wavelength division multiplexing}
\nAcronym{WSS}{\acroSCaps{wss}}{wavelength selective switch}

\nAcronym{3DWG}{\acroSCaps{3dwg}}{3D-waveguide}

%% file: table.tex
\def\arraystretch{0.9}
\begin{table}[t]
\caption{Simulation parameters.}
\begin{center}
\label{tab:syspar}
{\footnotesize
\begin{tabular}{lr}


\toprule
{\bf Parameter name} & {\bf Value}  \\
\midrule
WDM Channels & 11    \\
Symbol rate & \SI{41.79}{\giga \baud}   \\
Root-raised-cosine roll-off factor & \SI{1}{\percent} \\
Channel frequency spacing & \SI{50}{\giga \hertz}      \\
Center wavelength  & \SI{1550}{nm}  \\
Aggregate launch power & \SI{9.5}{\dBm}\\
Attenuation & \SI[per-mode=reciprocal]{0.2}{\decibel\per\km} \\
Dispersion parameter & \SI[per-mode=reciprocal]{17}{\ps\per\nm\per\km} \\
Nonlinearity parameter & \SI[per-mode=reciprocal]{1.3}{\per\W\per\km} \\
Fibre span length & 7\SI{5}{\km} \\
EDFA noise figure & \SI{5}{\dB} \\
\botrule
\end{tabular}
}
\end{center}
\end{table}